\input{aipcheck}
\documentclass[
    ,final            
  ]
  {aipproc}

\layoutstyle{6x9}

\begin{document}

\title{Spin Sum Rules and the Strong Coupling Constant at large distance.}

\classification{12.38Qk,11.55Hx}
\keywords      {Strong coupling constant, QCD spin sum rules, 
non-perturbative, commensurate scale relations, 
Schwinger-Dyson, Lattice QCD, AdS/CFT}

\author{A. Deur}{
  address={Thomas Jefferson National Accelerator Facility, 
Newport News, VA 23606}}

\begin{abstract}
We present recent results on the Bjorken and the
generalized forward spin polarizability sum rules from Jefferson Lab Hall A and CLAS
experiments, focusing on the low $Q^2$ part of the measurements. We then discuss the
comparison of these results with Chiral Perturbation theory calculations. In the 
second part of this paper, 
we show how the Bjorken sum rule with its connection to the
Gerasimov-Drell-Hearn sum, allows us to conveniently define an effective coupling
for the strong force at all distances.
\end{abstract}

\maketitle

\section{Introduction}
The information on the longitudinal spin structure of the nucleon is contained in the
$g_1(x,Q^2)$ and $g_2(x,Q^2)$ spin structure functions, with $Q^2$ the squared 
four-momentum transfered from the beam to the target, and $x=Q^2/(2M \nu)$ the
Bjorken scaling variable ($\nu$ is the energy transfer and $M$ the nucleon mass). 
The variable $Q^2$ indicates the space-time scale at which the nucleon is probed and
$x$ is interpreted in the parton model as the fraction of nucleon momentum carried by 
the struck quark.
 
Although  spin structure functions are the basic observables for nucleon spin studies, 
considering their integrals taken 
over $x$ is advantageous because of  
resulting simplifications. More importantly, such integrals are at the core
of the relation dispersion formalism. Relation dispersions relate the integral over
the imaginary part of a quantity to its real part. Expressing the 
imaginary part in function of the real part using the optical theorem yields 
\emph{sum rules}. When additional hypotheses are used, such as a low energy theorem
or the validity of Operator Product Expansion (OPE), the sum rules relate the integral 
to a static property of the target. If the static property is well known, the 
verification of the sum rule provides a check of the theory and hypotheses used in the 
sum rule derivation. When the property is not known because e.g. it is difficult to 
measure directly, sum rules can be used to access them. In that case, the theoretical
framework used to derived the sum rule is assumed to be valid. Details on integrals 
of spin structure functions and sum rules are given e.g. in the review~\cite{review moments}. 

Several spin sum rules exists. We
will focus on the Bjorken sum rule~\cite{Bjorken} and spin polarizability sum rules. We
will only briefly discuss the Gerasimov-Drell-Hearn (GDH) sum rule~\cite{GDH} since it is 
covered by V. Sulkosky's talk at this conference. In this paper, we will 
consider the $n$-th Cornwall-Norton moments: $\int_{0}^{1}dx g_{1}^{N}(x,Q^2) x^n$, with
$N$ standing for proton or neutron, and 
write the first moments as $\Gamma_{1}^{N}(Q^2)\equiv\int_{0}^{1}dx g_{1}^{N}(x,Q^2)$.

\section{The generalized Bjorken and GDH sum rules}
The Bjorken sum rule~\cite{Bjorken} relates the integral over $(g_1^p-g_1^n)$ 
to the nucleon axial charge $g_A$. This relation has been essential for understanding the 
nucleon spin structure and establishing, \emph{via} its 
$Q^2$-dependence, that Quantum Chromodynamics (QCD) describes 
the strong force when spin is included.
The Bjorken integral has been measured in polarized deep inelastic
lepton scattering (DIS) at SLAC, CERN and DESY~\cite{SLAC}-\cite{HERMES}
and at moderate $Q^2$ at Jefferson Lab (JLab)~\cite{EG1a/E94010}-\cite{RSS}. 
In the perturbative QCD (pQCD) domain (high 
$Q^2$) the sum rule reads:
\begin{eqnarray}
\label{eq:bj(Q2)}
\Gamma_{1}^{p-n}(Q^2)\equiv\int_{0}^{1}dx
\left( g_{1}^{p}(x,Q^2)-g_{1}^{n}(x,Q^2) \right)=
\hspace{1cm}\\
\frac{g_{A}}{6}\left[1-\frac{\alpha_{s}}{\pi}-3.58
\frac{\alpha_{s}^{2}}{\pi^{2}}-
20.21\frac{\alpha_{s}^{3}}{\pi^{3}}+...\right]+
\sum_{i=2}^{\infty}{\frac{\mu_{2i}^{p-n}(Q^{2})}{Q^{2i-2}}} \nonumber
\end{eqnarray}
where  $\alpha_s(Q^2)$ is the strong coupling strength. The bracket term (known 
as the leading twist term) is mildly dependent on $Q^2$ due
to pQCD soft gluon radiation. The other term contains non-perturbative 
power corrections (higher twists). These are quark and gluon 
correlations describing the nucleon structure
away from the large $Q^2$ (small distances) limit.

The generalized Bjorken sum rule has been derived for small distances. 
For large distances, in the $Q^2 \rightarrow 0$ limit, one finds the 
generalized GDH sum rule. The sum rule was first derived at $Q^2=0$:

\begin{equation}
\int_{\nu_{0}}^{\infty}\frac{\sigma_{1/2}(\nu)-\sigma_{3/2}(\nu)}
{\nu}d\nu=-\frac{2\pi^{2}\alpha\kappa^{2}}{M_t^{2}}\label{eq:gdh}
\end{equation}

where $\nu_{0}$ is the pion photoproduction threshold, $\sigma_{1/2}$ and 
$\sigma_{3/2}$ are the helicity dependent photoproduction cross sections 
for total photon plus target helicities 1/2 and 3/2, $\kappa$ is the anomalous
magnetic moment of the target while $S$ is its spin and $M_t$ its mass. $\alpha$
is the fine structure constant.

Replacing the photoproduction cross sections by the electroproduction ones 
generalized the left hand side of Eq.~\ref{eq:gdh} to any $Q^2$. Such generalization 
depends
on the choice of convention for the virtual photon flux, see e.g. 
ref.~\cite{review moments}. X. Ji and J. Osborne~\cite{ji01} showed that the 
sum \emph{rule} itself  (i.e. the whole Eq.~\ref{eq:gdh}) can be generalized as:

\begin{equation}
\frac{8}{Q^2}\int_0^{x^-} g_1dx=s_1(0,Q^2)\label{eq:gdh*}
\end{equation} 

where $S_1(\nu,Q^2)$ is the spin dependent Compton amplitude. This generalization
of the GDH sum rule makes the connection between the Bjorken and GDH generalized sum 
rules evident: GDH$=\frac{Q^2}{8} \times $Bjorken. 

The connection between the GDH and Bjorken sum rules allows us in principle to 
compute the moment $\Gamma_1$ at any $Q^2$. Thus, it provides us with a choice
observable to understand the transition of the strong force  from small to large distances.

\section{Experimental measurements of the first moments}

\begin{figure}
\includegraphics[scale=0.26]{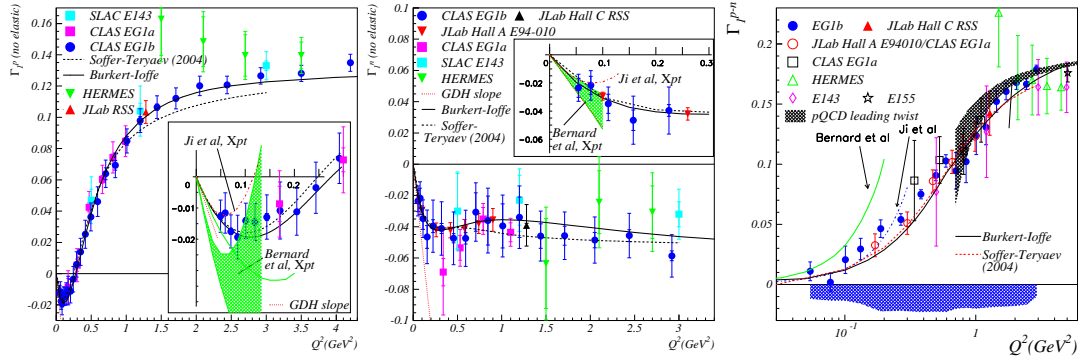}
\caption{\label{gammas} 
(Color online) Experimental data from SLAC, CERN, DESY and JLab at low and intermediate 
$Q^2$ on $\Gamma_1^p$ (left), $\Gamma_1^n$ (center) and $\Gamma_1^{p-n}$ (right).}
\end{figure}

Results from experimental measurements from SLAC~\cite{SLAC}, CERN~\cite{SMC}, 
DESY~\cite{HERMES} and JLab~\cite{eg1a proton}-\cite{RSS} of the first moments are shown in 
Figure~\ref{gammas}.
There is an excellent mapping of the moments at intermediate $Q^2$ and enough data points
a low $Q^2$ to start testing the Chiral Perturbation Theory ($\chi PT$), the effective 
theory strong force at large distances. In particular, the Bjorken sum is 
important for such test because the (p-n) subtraction cancels the $\Delta_{1232}$ 
resonance contribution which should make the $\chi PT$ calculations significantly more 
reliable~\cite{Burkert Delta}.
The comparison between the data at low $Q^2$ and $\chi PT$ 
calculations~\cite{meissner},\cite{Ji chipt}  can be seen 
more easily in the insert in each plot of Fig.~\ref{gammas}. The 
calculations assume the $\Gamma_1$ slope at $Q^2$=0 from the GDH sum rule prediction. 
Consequently, $\chi PT$ calculates the deviation from the slope and this is what one 
should test. A meaningful comparison is provided by fitting the lowest data points
using the form $\Gamma_1^N=\frac{\kappa_N^2}{8M^2}Q^2+aQ^4+bQ^6...$ and compare the
obtained value of $a$ to the values calculated from $\chi PT$. Such comparison has been 
carried out for the proton, deuteron~\cite{EG1b moments} and the Bjorken 
sum~\cite{Bj EG1b} (the results are given in Fig.~\ref{table}). These fits point out 
the importance of including a $Q^6$ 
term for $Q^2<0.1$ GeV$^2$. The $\chi PT$ calculations seems to agree best with the 
measurement of the Bjorken sum, in accordance with the discussion in~\cite{Burkert Delta}. 
Phenomenological models~\cite{AO},\cite{soffer} are in good agreement with the data
over the whole $Q^2$ range.

\section{Spin polarizability sum rules}
Higher moments of $g_1$ and $g_2$ are connected by sum rules to spin polarizabilities.
Those characterize  the coherent response of the 
nucleon to photons. They are defined using low-energy theorems in the form 
of a series expansion in the photon energy.  The first term of the 
series comes from the spatial distribution of charge and current (form 
factors) while the second term results from the deformation of these 
distributions induced by the photon (polarizabilities). Hence, 
polarizabilities are as important as form factors in understanding 
coherent nucleon structure. \emph{Generalized} spin polarizabilities 
describe the response to \emph{virtual} photons. The low energy theorem 
defining the generalized forward spin polarizability $\gamma_{0}$ is:
\begin{eqnarray}
\Re e[g_{TT}(\nu,Q^{2})-g_{TT}^{p\hat{o}le}(\nu,Q^{2})]= \label{eq:sr2} 
(\frac{2\alpha}{M^{2}})I_{TT}(Q^{2})\nu+\gamma_{o}(Q^{2})\nu^{3}+O(\nu^{5}),
\end{eqnarray}
where $g_{TT}$ is the spin-flip doubly-virtual Compton scattering 
amplitude, and $I_{TT}$ is the coefficient of the $O(\nu)$ term of the 
Compton amplitude which can be used to generalize the 
GDH sum rule to non-zero $Q^2$. We have $I_{TT}(Q^{2}=0)=\kappa /4$. In 
practice $\gamma_{0}$ can be obtained from a sum rule which has a 
derivation akin to that of the GDH sum rule:
\begin{eqnarray}
\gamma_{0}=\frac{16\alpha M^{2}}{Q^{6}}\int_{0}^{x_{0}}x^{2}\left(g_{1}-
\frac{4M^{2}}{Q^{2}}x^{2}g_{2}\right)dx, 
\label{eq:srg0}
\end{eqnarray}
Similar relations define the generalized longitudinal-transverse polarizability 
$\delta_{LT}$ :
\begin{eqnarray}
\Re e[g_{LT}(\nu,Q^{2})-g_{LT}^{p\hat{o}le}(\nu,Q^{2})]=
(\frac{2\alpha}{M^{2}})QI_{LT}(Q^{2})+Q\delta_{LT}(Q^{2})\nu^{2}+O(\nu^{4}),
\end{eqnarray}
\begin{eqnarray}
\delta_{LT} = \frac{16\alpha M^{2}}{Q^{6}}\int_{0}^{x_{0}}x^2\left( g_{1}+g_{2} \right)dx.\label{eq:srdlt}
\end{eqnarray}
where $g_{LT}$ is the longitudinal-transverse interference
amplitude, and $I_{LT}$ is the coefficient of the $O(\nu)$ term of the 
Compton amplitude. Details on the derivation of 
Eqs.~\ref{eq:sr2}-\ref{eq:srdlt} can be found in~\cite{review moments}. Higher 
moments are advantageous because they are essentially free of the 
uncertainty associated with the low-$x$ extrapolation necessary since reaching $x \rightarrow 0$ would require an infinite beam energy. Eqs.~\ref{eq:srg0} and \ref{eq:srdlt}
are examples of uses of sum rules to measure observables that are otherwise hard
to access.

\begin{figure}
\includegraphics[scale=0.24]{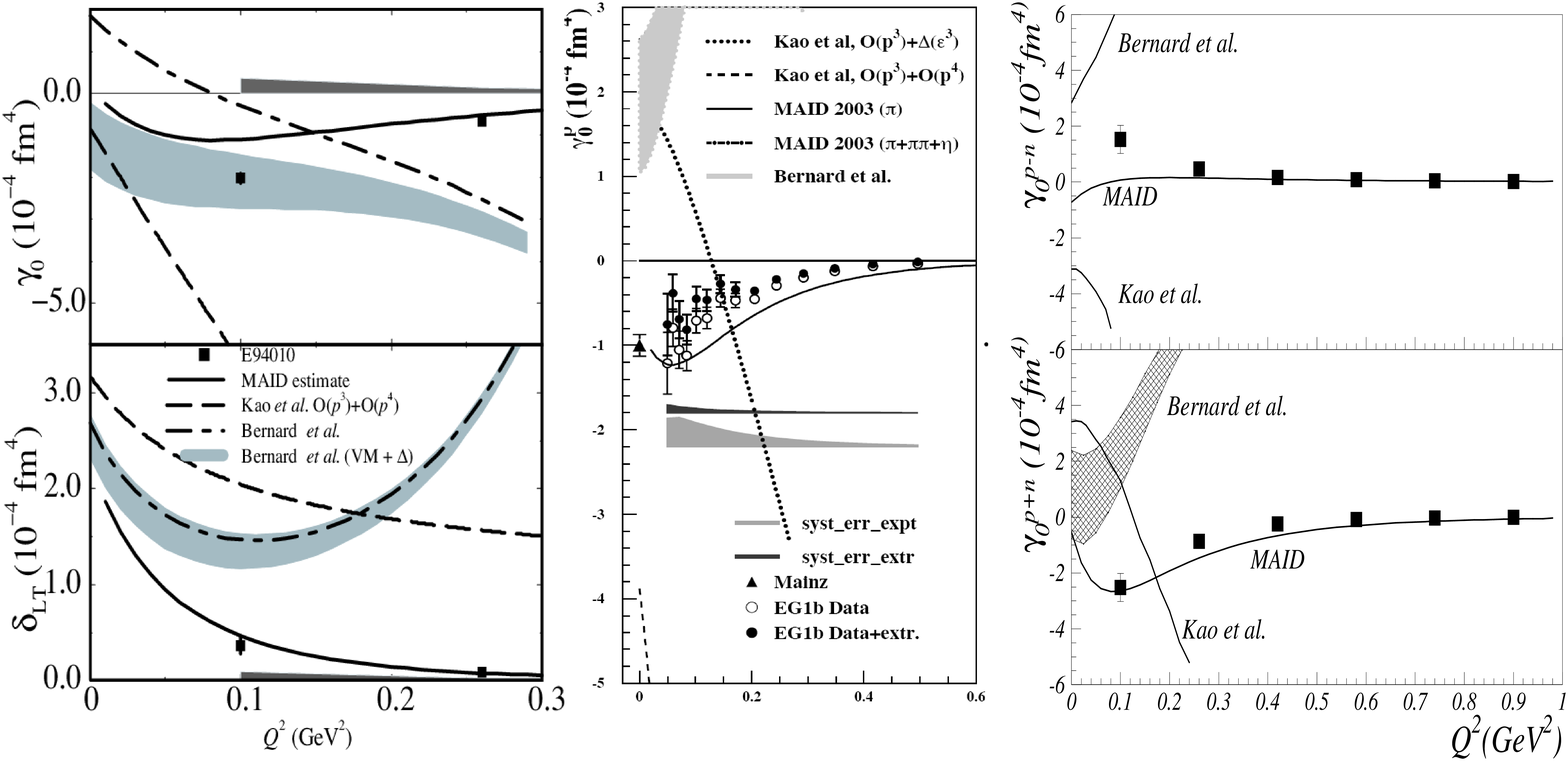}
\caption{\label{gamma0s} 
Experimental data at low $Q^2$ on generalized spin polarizabilities. Results on neutron
(JLab Hall A experiment E94010~\cite{E94010-3}) are shown on the left 
(top: $\gamma_0^n$, bottom: $\delta_{LT}^{n}$). Results on the proton (JLab CLAS 
experiment EG1b~\cite{EG1b moments}) $\gamma_0^p$ are shown in the central plot. The isospin decomposition
of $\gamma_0$ (E94010+EG1b~\cite{Bj EG1b}) is shown on the right (top: $\gamma_0^{p-n}$, bottom 
$\gamma_0^{p+n}$.}
\end{figure}

\begin{figure}
\includegraphics[scale=0.26]{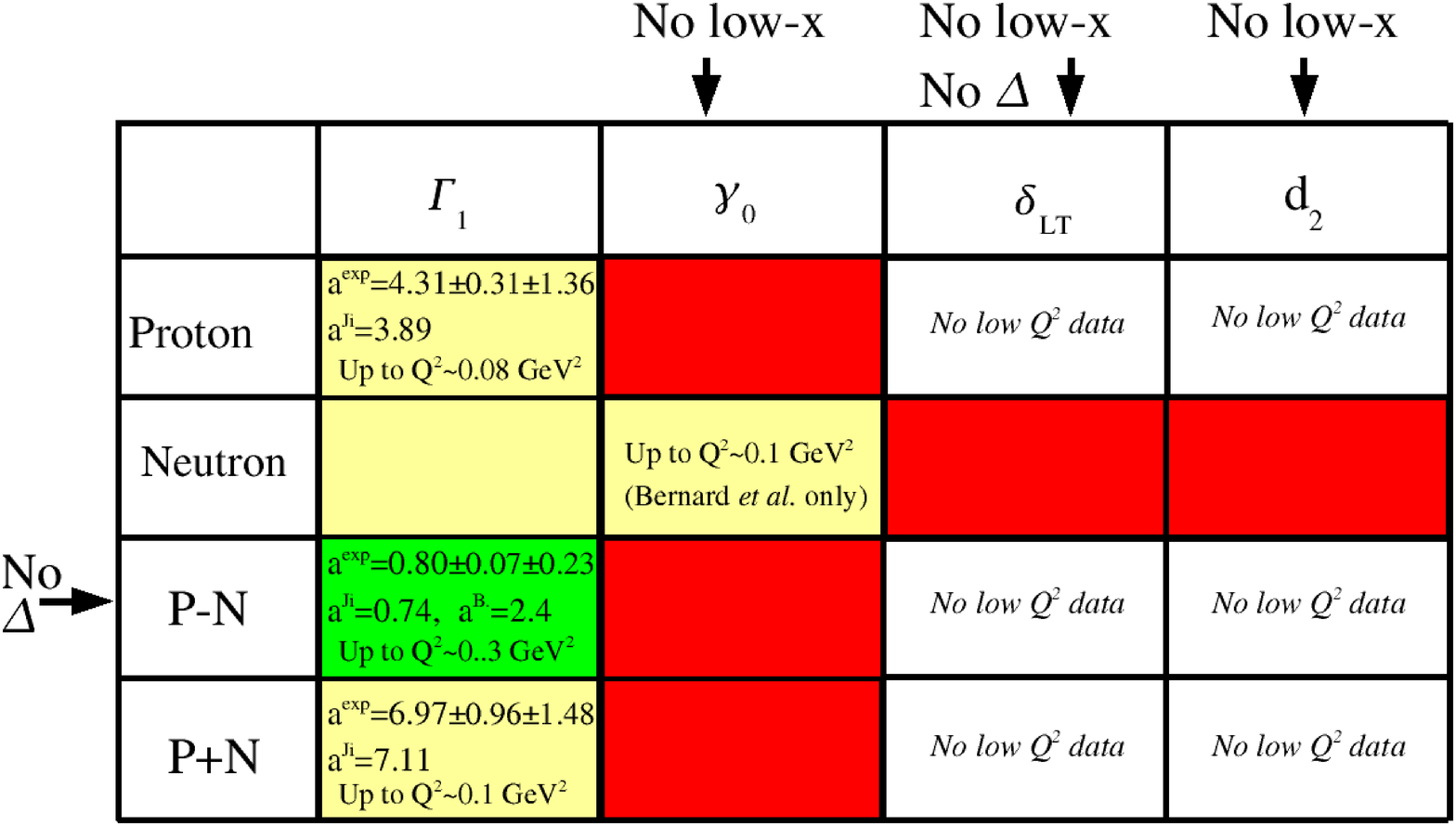}
\caption{\label{table} 
(Color online) Summary the comparison between $\chi PT$ calculations and data. The green indicates
a good match within the region in which we expect the chiral perturbation series to 
be reliable, 
the yellow an agreement over a shorter $Q^2$ range, and the red a mismatch. }
\end{figure}

In the case of the transverse-longitudinal polarizability $\delta_{LT}$, 
the $\Delta_{1232}$ contribution
is suppressed at low $Q^2$ because the N-$\Delta$ transition is mostly
transverse, making the contribution of the $\Delta$ to the longitudinal-transverse 
(LT) interference term  very small. Thus $\delta_{LT}$ should also provide a reliable 
test of $\chi PT$ computations. Furthermore, as for the Bjorken sum 
the isovector part 
of $\gamma_{0}$, $\gamma_{0}^{p}-\gamma_{0}^{n}$, should offers similar advantages 
for checking the calculation techniques of $\chi PT$. 
The low $Q^2$ data on forward spin polarizabilities, from Hall A E94010 and CLAS EG1b, 
are shown on Fig.~\ref{gamma0s}
There is no agreement between the data and the $\chi PT$ calculations (except possibly
for the lowest $Q^2$ point of $\gamma_0^n$ that agrees with the explicitly
covariant calculation of Bernard \emph{et al}). Such disagreement is surprising 
because the first point should be into the validity domains of $\chi PT$.
It is even more surprising for 
$\delta_{LT}^{n}$ and $\gamma_0^{p-n}$ because of the $\Delta$ suppression for these
two quantities. This points out that including the $\Delta$ in the calculations may not 
be the only challenge facing $\chi PT$ theorists. In contrast, the MAID model~\cite{MAID}
is in good agreement with the data. 
Figure~\ref{table} summarizes the comparison between $\chi PT$ calculations and data.
(We added the higher moment $d_2^n$ measured in Experiment E94010). 

\section{The strong coupling at large distances}
So far, we have discussed the data at low $Q^2$. The primary goal of the JLab experiments
was to map precisely the intermediate $Q^2$ range in order to shed light on the 
transition from short distances (where the degrees of freedom pertinent to the strong force
are the partonic ones) to large distances where the hadronic degrees of freedom are 
relevant to the strong force. One feature seen on Fig.~\ref{gammas} is that the transition
from small to large distances is smooth, e.g. without sign of a phase transition. This
fact can be used to extrapolate the definition of the strong force effective coupling
to large distances. Before discussing this, we first review the QCD coupling 
and the issues with calculating it at large distances.

In QCD, the magnitude of the strong force is given by 
the running coupling constant $\alpha_{s}$. 
At large $Q^2$, in the pQCD domain, $\alpha_{s}$ is well defined and 
is given by the series:
\vspace{-0.2cm}
\begin{eqnarray}
\mu\frac{\partial\alpha_{s}}{\partial\mu} & =2\beta(\alpha_{s}) & 
=-\frac{\beta_{0}}{2\pi}\alpha_{s}^{2}-\frac{\beta_{1}}{4\pi^{2}}
\alpha_{s}^{3}-\frac{\beta_{2}}{64\pi^{3}}\alpha_{s}^{4}-...
\label{eq:alpha_s beta serie}
\end{eqnarray}
Where $\mu$ is the energy scale, to be identified to $Q$. The 
first terms of the $\beta$ series are: 
$\beta_{0}=11-\frac{2}{3}n$ with $n$ the number of active quark 
flavors, $\beta_{1}=51-\frac{19}{3}n$ and $\beta_{2}=2857-
\frac{5033}{9}n+\frac{325}{27}n^{2}$.
The solution of the differential equation \ref{eq:alpha_s beta serie} is:
\small
\vspace{-0.2cm}
\begin{eqnarray}
\alpha_{s}(\mu)=\frac{4\pi}{\beta_{0}ln(\mu^{2}/\Lambda_{QCD}^{2})}
\times \label{eq:alpha_s}\\
 &  & \hspace{-5cm} \left[1-\frac{2\beta_{1}}{\beta_{0}^{2}}
\frac{ln\left[ln(\mu^{2}/\Lambda_{QCD}^{2})\right]}{ln(\mu^{2}/
\Lambda_{QCD}^{2})}+\frac{4\beta_{1}^{2}}{\beta_{0}^{4}ln^{2}
(\mu^{2}/\Lambda_{QCD}^{2})}\left(\left(ln\left[ln(\mu^{2}/
\Lambda_{QCD}^{2})\right]-\frac{1}{2}\right)^{2}+\frac{
\beta_{2}\beta_{0}}{8\beta_{1}^{2}}-\frac{5}{4}\right)\right]
\nonumber 
\end{eqnarray}
\normalsize

\noindent Equation \ref{eq:alpha_s} allows us to evolve the different experimental 
determinations of $\alpha_{s}$ to a conventional scale, typically 
$M_{z_{0}}^{2}$. 
The agreement between the $\alpha_{s}$ obtained from different observables
demonstrates its universality and the validity of 
Eq. \ref{eq:alpha_s beta serie}. One can obtain  
$\alpha_{s}(M_{z_{0}}^{2})$ with doubly polarized DIS data 
and assuming the validity of the Bjorken sum. Solving 
Eq. \ref{eq:bj(Q2)} using the experimental value of 
$\Gamma_{1}^{p-n}$, and then using Eq. \ref{eq:alpha_s} provides 
$\alpha_{s}(M_{z_{0}}^{2})$.

Equation \ref{eq:alpha_s} leads to an infinite coupling at large distances, 
when $Q^2$ approaches $\Lambda^{2}_{QCD}$. 
This is not a conceptual problem since we are out of the validity 
domain of pQCD on which Eq. \ref{eq:alpha_s} is based.
But since data show no sign of discontinuity or phase transition 
when crossing the intermediate $Q^{2}$ domain, one 
should be able to define an effective coupling $\alpha_{s}^{eff}$ at 
any $Q^2$ that matches $\alpha_{s}$ at large $Q^{2}$ but stays finite at 
small $Q^{2}$. 

The Bjorken Sum Rule can be used to define $\alpha_{s}^{eff}$ at 
low Q$^{2}$. Defining $\alpha_{s}^{eff}$
from a pQCD equation truncated to first order (in our case Eq. (\ref{eq:bj(Q2)}: 
$\Gamma_{1}^{p-n}\equiv\frac{1}{6}(1-\alpha_{s,g_{1}}/\pi)$),
offers advantages. In particular, $\alpha_{s}^{eff}$ does not 
diverge near $\Lambda_{QCD}$ and is renormalization scheme independent. 
However, $\alpha_{s}^{eff}$ becomes dependent 
on the choice of observable employed to define it. If $\Gamma_{1}^{p-n}$ is 
used as the defining observable, the effective coupling is noted 
$\alpha_{s,g_{1}}$.  
Relations, called \emph{commensurate scale relations} \cite{Brodsky CSR}, 
link the different effective couplings so in principle one
effective coupling is enough to describe the strong force and the theory 
retains its predictive power. These relations are defined for short distances
and whether they extrapolate to large distances remains to be investigated. 

The choice of defining the effective charge with the Bjorken sum has many advantages: 
low $Q^2$ data exist and near real photons data from JLab is being 
analyzed~\cite{gdh neutron,gdh proton}. Furthermore, sum rules constrain
$\alpha_{s,g_1}$ at both low and large $Q^2$, as will be discussed in
the next paragraph. Another advantage is that, as discussed for the low $Q^2$ 
domain, the simplification arising in $\Gamma _1^{p-n}$ makes a quantity 
well suited to be calculated at any $Q^2$~\cite{Burkert Delta}. These simplifications 
are manifest 
at large $Q^2$ when comparing the validities of the Bjorken and Ellis-Jaffe sum rules.
It also simplifies Lattice QCD calculations in the intermediate $Q^2$ domain. 
Finally, it may be argued that $\alpha_{s,g_1}$ might be more directly 
comparable to theoretical calculations than other effective couplings extracted from other
observables: part of the coherent response of the nucleon
is suppressed in the Bjorken sum, e.g. the $\Delta$ resonance, so the non-resonant
background, akin to the pQCD incoherent scattering process, contributes especially
importantly to the Bjorken sum. This argument is reinforced if global duality works, 
a credible proposal since the  $\Delta$ resonance is suppressed.

\begin{figure}
\includegraphics[scale=0.45]{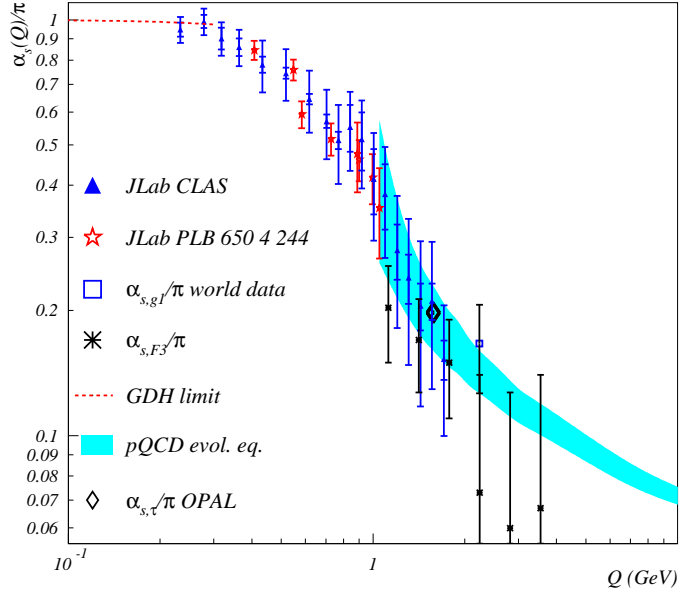}
\caption{\label{fig: alpha_s eff} 
Value of $\alpha_{s,g_{1}}/\pi$ 
extracted from the world data on the Bjorken sum at 
$Q^{2}=5$ GeV$^{2}$ \cite{E155-E155x} and from JLab 
data~\cite{eg1a proton,Bj EG1b}. Also shown are 
$\alpha_{s,\tau}$ extracted from the OPAL 
data on $\tau$ decay \cite{Brodsky CSR}, and $\alpha_{s,GLS}$ extracted 
using the 
Gross-Llewellyn Smith sum rule \cite{GLS} and its measurement by the 
CCFR collaboration \cite{CCFR}.
The gray band indicates $\alpha_{s,g_{1}}$ extracted from the pQCD expression 
of the Bjorken sum at leading twist and third order in $\alpha_{s}$ (with $\alpha_{s}$ 
computed using Eq. \ref{eq:alpha_s}). The values of 
$\alpha_{s,g_{1}}/\pi$ extracted using the Gerasimov-Drell-Hearn  
sum rule is given by the red dashed line.} 
\end{figure}

\begin{figure}
\vspace{-0.8cm}
\includegraphics[scale=0.55]{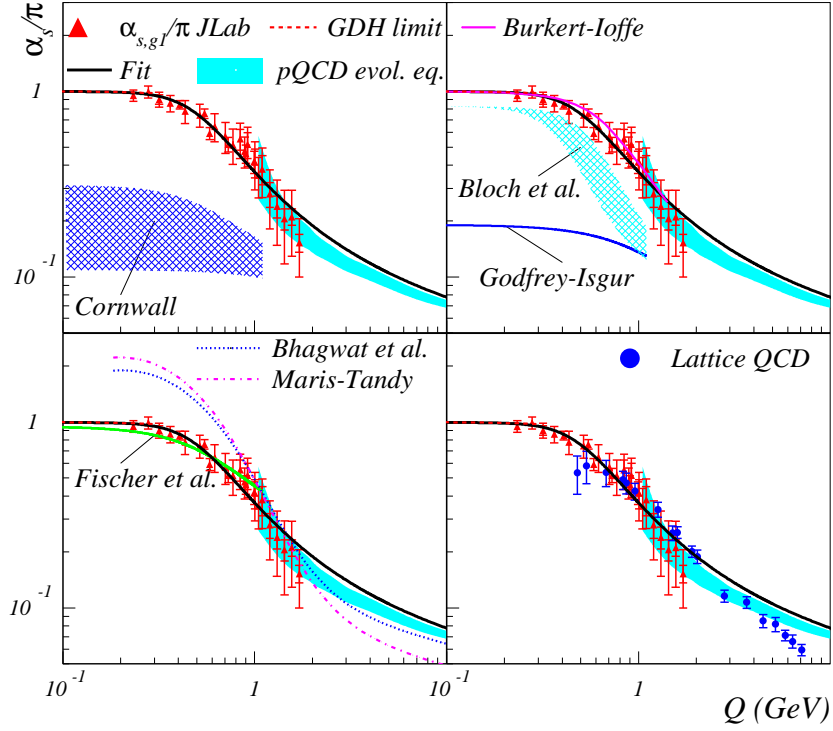}
\caption{\label{fig: alpha_s eff4} The effective coupling 
$\alpha_{s,g_{1}}$ extracted from JLab data, its fit, and its 
extraction using the Burkert and Ioffe \cite{AO} model to 
obtain $\Gamma_{1}^{p-n}$. The $\alpha_{s}$ calculations are: 
Top left: Schwinger-Dyson equations (Cornwall \cite{Cornwall}); 
Top right: Schwinger-Dyson equations (Bloch) \cite{Bloch} and 
$\alpha_{s}$ used in a quark constituent model \cite{Godfrey-Isgur};
Bottom left: Schwinger-Dyson equations (Maris-Tandy \cite{Tandy}),
Fischer, Alkofer, Reinhardt and Von Smekal \cite{Fischer} and Bhagwat 
et al. \cite{Bhagwat}; Bottom right: Lattice QCD \cite{Furui}.
}
\end{figure}

The effective coupling definition in terms of 
pQCD evolution equations truncated to first order was proposed by 
Grunberg \cite{Grunberg}. Grunberg's definition is meant for
short distances but one can always extrapolated this definition and see 
how the resulting coupling compares to calculation
of $\alpha_{s}$ at large distances. 
Using Grunberg's definition at large distances entails
including higher twists in $\alpha_{s,g_{1}}$ in addition to the higher terms of the
pQCD series. Effective couplings have been extracted from different 
observables and have been compared to each other
using the commensurate scale relations~\cite{deur alpha_s^eff}, see 
Fig. \ref{fig: alpha_s eff}. 
There is good agreement between $\alpha_{s,g_{1}}$, $\alpha_{s,F_{3}}$ and
 $\alpha_{s,\tau}$. The GDH and Bjorken sum rules can be used to extract  
$\alpha_{s,g_{1}}$ at small and large $Q^{2}$ respectively 
\cite{deur alpha_s^eff}. This, together with the JLab data at 
intermediate $Q^{2}$, provides for the first time a coupling 
at any $Q^{2}$. A striking feature of Fig. \ref{fig: alpha_s eff} 
is that $\alpha_{s,g_{1}}$ becomes scale invariant at small $Q^{2}$. 
This was predicted by a number of calculations and it is 
known that color confinement leads to an
infrared fixed point \cite{irfp}, but it is the 
first time it is seen experimentally.
A fit of the $\alpha_{s,g_{1}}$ has been 
performed and is shown on Fig. \ref{fig: alpha_s eff4} 
(plain black line). 

In Figure \ref{fig: alpha_s eff4}, $\alpha_{s,g_{1}}$ is compared to 
theoretical results. There are several techniques used to predict 
$\alpha_{s}$ at small $Q^{2}$, e.g. lattice QCD, solving the 
Schwinger-Dyson equations, or choosing the coupling in a constituent 
quark model so that it reproduces hadron spectroscopy. However, the 
connection 
between  these $\alpha_{s}$ is unclear, in part because of the different 
approximations used. In addition, the precise relation 
between $\alpha_{s,g_{1}}$ (or any effective coupling defined using 
\cite{Grunberg} or \cite{Brodsky CSR}) and these computations is
unknown. Nevertheless, one can still compare them to see 
if they share common features. The calculations and $\alpha_{s,g_{1}}$ 
present 
a similar behavior. Some calculations, in
particular the lattice one, are in excellent agreement with 
$\alpha_{s,g_{1}}$. 

These works show that $\alpha_{s}$ is scale invariant (\emph{conformal 
behavior}) at small 
and large $Q^{2}$ (but not in the transition region between the 
fundamental description of QCD in terms of quarks and gluons 
degrees of freedom and its effective one in terms of baryons and mesons).
The scale invariance at large $Q^2$ is the well known asymptotic freedom. 
The conformal behavior at small $Q^{2}$ 
is essential to apply a property of \emph{conformal field theories} 
(CFT) to the study of hadrons: the \emph{Anti-de-Sitter 
space/Conformal Field Theory (AdS/CFT) correspondence} of Maldacena 
\cite{Maldacena}, that links a strongly coupled gauge field to weakly 
coupled superstrings states. Perturbative calculations are feasible in 
the weak coupling AdS theory. They are then projected on the AdS 
boundary, where they correspond to the calculations that would have
been obtained with the strongly coupled CFT. This opens the possibility 
of analytic non-perturbative QCD calculations \cite{ads/CFT}.

\section{Summary and perspectives}
We discussed the data on moments of spin structure functions are large distances
and compared them to $\chi PT$, the strong force effective theory at large 
distances. The data and calculations do not consistently agree. In particular,
the better agreement expected for observables in which the $\Delta$ resonance
is suppressed is seen only for the Bjorken sum, but not for $\delta_{LT}^n$ or
 $\gamma_{0}^{p}-\gamma_{0}^{n}$. Apparently, the $\Delta$ cannot explained
single-handedly the discrepancy between data and calculations. The data shown 
were taken at JLab during experiments focusing on covering the intermediate $Q^2$ 
range~\cite{eg1a proton,Bj EG1b}. A new generation of 
experiments~\cite{gdh neutron,gdh proton} especially dedicated to push such
measurements to lower $Q^2$ and higher precision has taken new data that are 
being analyzed. In addition, a new experiment to measure $\delta_{LT}^p$ in Hall A 
at low $Q^2$
is approved~\cite{delta_lt_p}, while the frozen spin HD target recently arrived at 
JLab from BNL is opening new possibilities of measurements with CLAS using 
transversely polarized protons or deuterons. 

The smoothness of $Q^2$-dependence of the moments when transiting from perturbative to 
the non-perturbative domain allows to extrapolate the definitions of effective strong 
couplings from short to large distances. Thanks to the data on nucleon spin structure 
and to spin sum rules, the effective strong coupling $\alpha_{s,g_1}$can be extracted 
in any regime of QCD. The question of comparing it with theoretical calculations of
$\alpha_{s}$ at low $Q^2$ is open, but such comparison exposes a
similarity between these couplings. Apart for the parton-hadron transition 
region, the coupling shows
that QCD is approximately a conformal theory. This is a necessary
ingredient to the application of the AdS/CFT correspondence that may 
make analytical calculations possible in the non-perturbative domain of QCD.

\begin{theacknowledgments}This work is supported by the U.S. 
Department of Energy (DOE). The Jefferson Science Associates (JSA) 
operates the Thomas Jefferson National Accelerator Facility for the DOE 
under contract DE-AC05-84ER40150.  
\end{theacknowledgments}


\IfFileExists{\jobname.bbl}{}
 {\typeout{}
  \typeout{******************************************}
  \typeout{** Please run "bibtex \jobname" to optain}
  \typeout{** the bibliography and then re-run LaTeX}
  \typeout{** twice to fix the references!}
  \typeout{******************************************}
  \typeout{}
 }

\end{document}